\documentclass{PoS}

\newcommand{\mr}[1]{\ensuremath{\mathrm{#1}}}
\newcommand{\mi}[1]{\ensuremath{\mathit{#1}}}
\newcommand{\mb}[1]{\ensuremath{\mathbf{#1}}}
\newcommand{\ux}[2]{\mr{^{#1} #2}}

\title{\mb{^{44}Ti} and \mb{^{56}Ni} in core-collapse
supernovae}

\ShortTitle{$\mi{^{44}Ti}$ and $\mi{^{56}Ni}$ in core-collapse
supernovae}

\author{
\speaker{Georgios Magkotsios}$^{ab}$, Francis X. Timmes$^{abc}$,
Michael Wiescher$^{b}$, Christopher L. Fryer$^{ad}$, Aimee
Hungerford$^{ad}$, Patrick Young$^{ac}$, Michael E. Bennett$^{ae}$,
Steven Diehl$^{adf}$, Falk Herwig$^{aeg}$, Raphael Hirschi$^{ae}$,
Marco Pignatari$^{abe}$ and Gabriel Rockefeller$^{ad}$\\
\llap{$^a$}The NuGrid Collaboration\\
\llap{$^b$}Joint Institute for Nuclear Astrophysics, University of Notre Dame, IN, 46556, USA\\
\llap{$^c$}School of Earth and Space Exploration, Arizona State University, Tempe, AZ 85287, USA\\
\llap{$^d$}Computational Methods (CCS-2), Los Alamos National Laboratory, Los Alamos, NM, 87544, USA\\
\llap{$^e$}Astrophysics Group, Keele University, ST5 5BG, UK\\
\llap{$^f$}Theoretical Astrophysics Group (T-6), Los Alamos National Laboratory, Los Alamos, NM, 87544, USA\\
\llap{$^g$}Dept. of Physics \& Astronomy, Victoria, BC, V8W 3P6, Canada\\
E-mail:\email{gmagkots@nd.edu} }

\abstract{
We investigate the physical conditions where $\ux{44}{Ti}$ and
$\ux{56}{Ni}$ are created in core-collapse supernovae. In this preliminary
work we
use a series of post-processing network calculations with
parameterized expansion profiles that are representative of the wide
range of temperatures, densities and electron-to-baryon ratios found
in 3D supernova simulations.  Critical flows that affect the final
yields of $\ux{44}{Ti}$ and $\ux{56}{Ni}$ are assessed.
}

\FullConference{10th Symposium on Nuclei in the Cosmos\\
         July 27 - August 1 2008\\
         Mackinac Island, Michigan, USA}

\begin{document}

\section{Understanding the mechanisms that produce \mb{^{44}Ti}}

Cassiopeia A is still the only supernova remnant with an unambiguous
detection of \ux{44}{Ti} \cite{iyudinI, iyudinII, vink, ti44obs}.
Various arguments have been suggested to account for the apparent
paucity of such detections. In this work we examine if thermodynamic
pathways and nuclear physics responsible for the creation and
annihilation of \ux{44}{Ti} can be one of the reasons why there are
so few gamma-ray line detections of \ux{44}{Ti}.

\begin{figure}[ht]
\begin{center}
\includegraphics[width=0.60\textwidth]{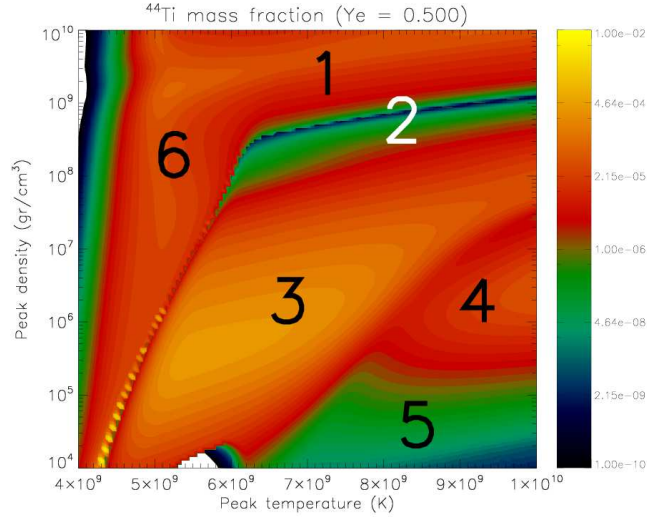}
\caption{Contour plot of \ux{44}{Ti} final yields as a function of
temperature and density for adiabatic expansions. The numbers
signify the various regimes for \ux{44}{Ti} nucleosynthesis.}
\label{fig:ye05_AD1}
\end{center}
\end{figure}

The temperature-density plane of interest for \ux{44}{Ti} is shown
in Fig.\ \ref{fig:ye05_AD1}.  Each point in the plane represents the
initial temperature and density of a material initially composed
with an electron fraction of \mr{Y_{e}}=0.5.  Each point in the
plane was then evolved along an adiabatic expansion profile under
the assumption of a constant radiation entropy \mr{T^3/\rho \sim}
constant. Colors in the plot correspond to the final \ux{44}{Ti}
yields, after adiabatic expansion has cooled the material to the
point where nuclear reactions cease.  The numbered regions
correspond to different evolutionary mechanisms.  Of particular
interest is the ``chasm'' of region 2 where very little \ux{44}{Ti}
is produced. For each region, only certain reactions affect the
yields of \ux{44}{Ti} and \ux{56}{Ni}:
\begin{enumerate}
\item Fast freeze-out from NSE. Abundance largely determined from Q-values.
\item Chasm region: Passage from NSE to QSE. Reactions:
\mr{\ux{44}{Ti}(\alpha,p)\ux{47}{V}},
\mr{\ux{20}{Ne}(\alpha,p)\ux{23}{Na}}, \\
\mr{\ux{21}{Na}(\alpha,p)\ux{24}{Mg}},
\mr{\ux{20}{Ne}(p,\gamma)\ux{21}{Na}}
\item Normal $\alpha$-rich freeze-out: \ux{56}{Ni} dominates.
Reactions: \mr{3\alpha\rightarrow\ux{12}{C}},
\mr{\ux{44}{Ti}(\alpha,p)\ux{47}{V}}, \\
\mr{\ux{20}{Ne}(\alpha,p)\ux{23}{Na}},
\mr{\ux{21}{Na}(\alpha,p)\ux{24}{Mg}},
\mr{\ux{45}{V}(p,\gamma)\ux{46}{Cr}},
\mr{\ux{57}{Ni}(p,\gamma)\ux{58}{Cu}}
\item $\alpha$- and $p$-rich freeze-out. Reactions:
\mr{3\alpha\rightarrow\ux{12}{C}},
\mr{\ux{45}{V}(p,\gamma)\ux{46}{Cr}},
\mr{\ux{44}{Ti}(p,\gamma)\ux{45}{V}}, \\
\mr{\ux{41}{Sc}(p,\gamma)\ux{42}{Ti}},
\mr{\ux{43}{Sc}(p,\gamma)\ux{44}{Ti}},
\mr{\ux{40}{Ca}(p,\gamma)\ux{41}{Sc}},
\mr{\ux{40}{Ca}(\alpha,p)\ux{43}{Sc}}
\item Photodisintegration regime: n, p and $\alpha$ dominate
\item Incomplete silicon burning: \ux{28}{Si} rich
\end{enumerate}
The importance of these reactions was determined from analysis of
the flows in a sensitivity survey.

\section{Chasm shift between expansion profiles for symmetric matter}

\begin{figure}[ht]
\begin{minipage}{0.5\textwidth}
\begin{center}
\includegraphics[width=\textwidth]{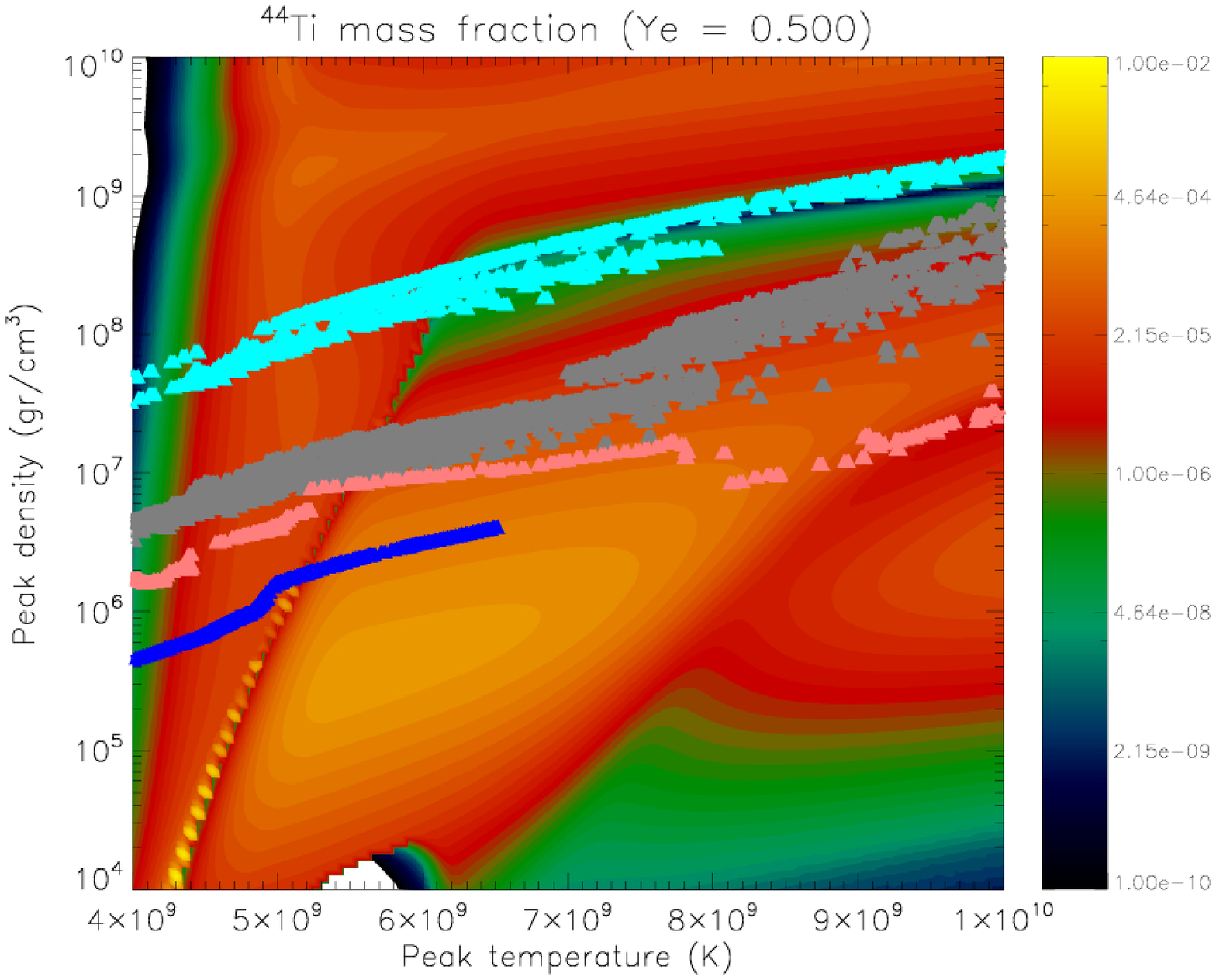}
\end{center}
\end{minipage}
\begin{minipage}{0.5\textwidth}
\begin{center}
\includegraphics[width=\textwidth]{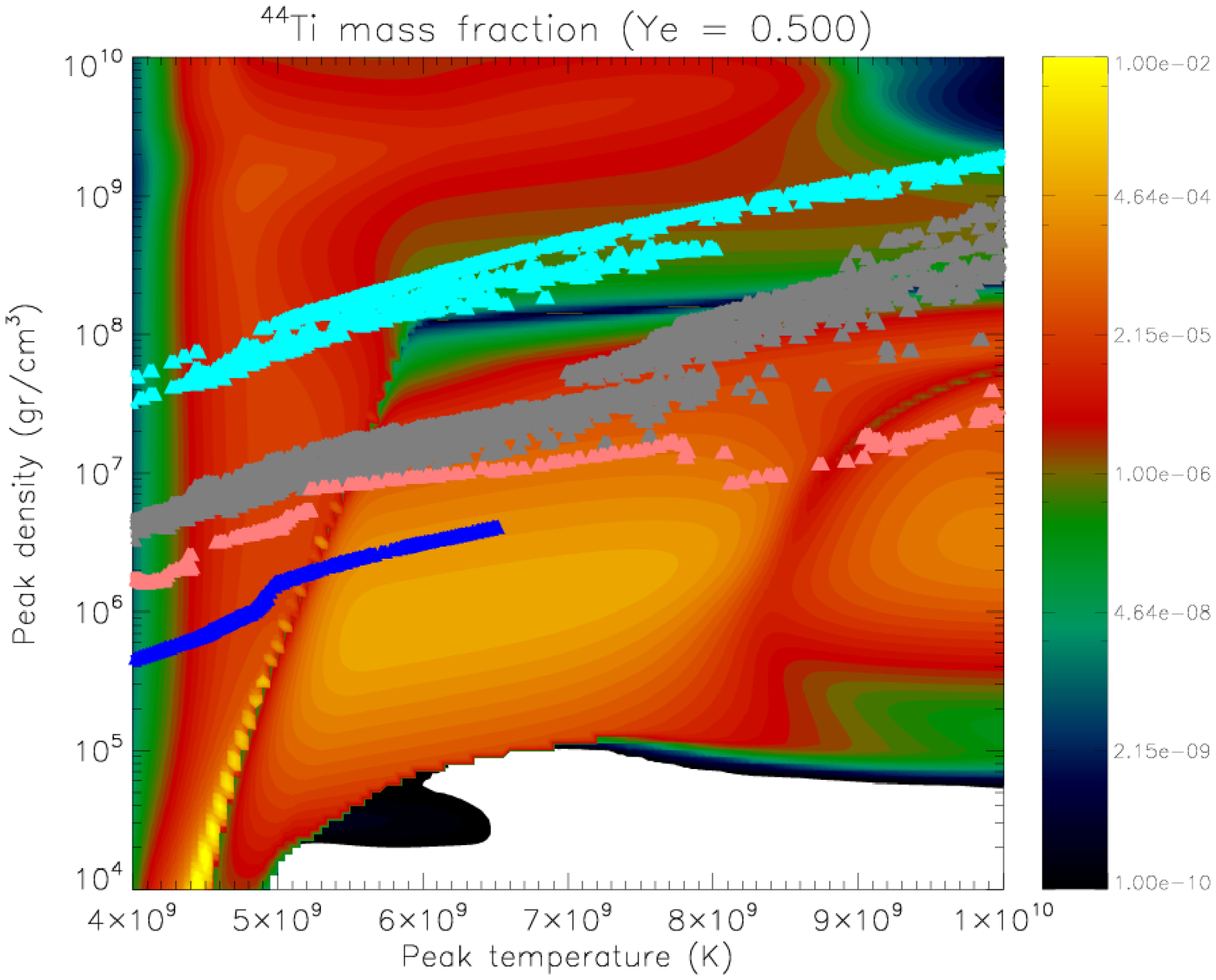}
\end{center}
\end{minipage}
\caption{$\ux{44}{Ti}$ contour plots for two expansion profiles. The
contour plot to the left is for an adiabatic expansion, while the
one to the right is for a power-law expansion. The embedded points
correspond to conditions met in supernovae simulations. The cyan
ones are based on a rotating MHD star, the grey on a rotating 2D
explosion model, the pink are for a gamma-ray burst model and the
blue ones on a model specifically for Cassiopeia A.}
\label{fig:ti44_AD1_PL2}
\end{figure}

Next we consider the effect of two different parameterizations on
how material cools down, the adiabatic expansion pathway described
above (left image in Fig.\ \ref{fig:ti44_AD1_PL2}) and a power-law
profile derived from 3D simulations of the supernova explosion
(right image in Fig.\ \ref{fig:ti44_AD1_PL2}). In general, for any
given $\mr{Y_{e}}$, a power-law expansion profile shifts the
\ux{44}{Ti} chasm to lower densities. Points from multi-dimensional
supernova simulations are overlayed on these two \ux{44}{Ti} maps,
and suggest that core-collapse supernovae may populate the region of
parameter space where very little \ux{44}{Ti} is produced.

\begin{figure}[ht]
\begin{center}
\includegraphics[width=0.60\textwidth]{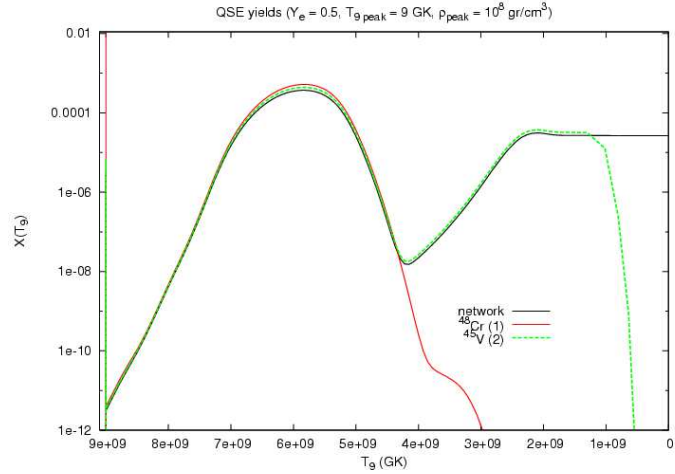}
\caption{$\ux{44}{Ti}$ mass fraction as a function of temperature.
The black curve is the outcome of a network code for a typical
$\alpha$-rich freeze-out. The red curve assumes that $\ux{44}{Ti}$
belongs to the local cluster of the \mr{N=22} and \mr{N=24}
isotones, connected by \mr{(\alpha,p)} reactions. $\ux{48}{Cr}$ is
the reference isotope. The green curve assumes that this equilibrium
between the isotones has broken, and $\ux{44}{Ti}$ is in equilibrium
only with certain isotopes along the \mr{N=22} isotone. $\ux{45}{V}$
is the reference isotope in this case.} \label{fig:ti44_qse_temp}
\end{center}
\end{figure}

The reaction that accounts for most of the chasm's location is
\mr{\ux{44}{Ti}(\alpha,p)\ux{47}{V}}. This reaction controls a
small-scale equilibrium cycle which keeps \ux{44}{Ti} in the
large-scale silicon group QSE cluster. The reaction's equilibrium
persistence decreases the \ux{44}{Ti} mass fraction (Fig.\
\ref{fig:ti44_qse_temp}).  For appropriate initial conditions, the
expansion timescale is slow enough that this equilibrium persists
until the end of nucleosynthesis. For these conditions, \ux{44}{Ti}
will be depleted and a chasm appears.

The \mr{\ux{44}{Ti}(\alpha,p)\ux{47}{V}} goes out of its equilibrium
cycle at a threshold temperature of \mr{T_{9}\sim 3} GK.  Since the
timescales for different expansion profiles are different, the
density at the point in time when this threshold temperature is
reached will differ. Different densities when this threshold
temperature is reached imply different initial densities for the
same initial temperature. The chasm is thus located at lower initial
densities for longer expansion timescales.

\section{Calculations for asymmetric matter}

\begin{center}
\begin{figure}[ht]
\begin{minipage}{0.33\textwidth}
\begin{center}
\includegraphics[width=\textwidth]{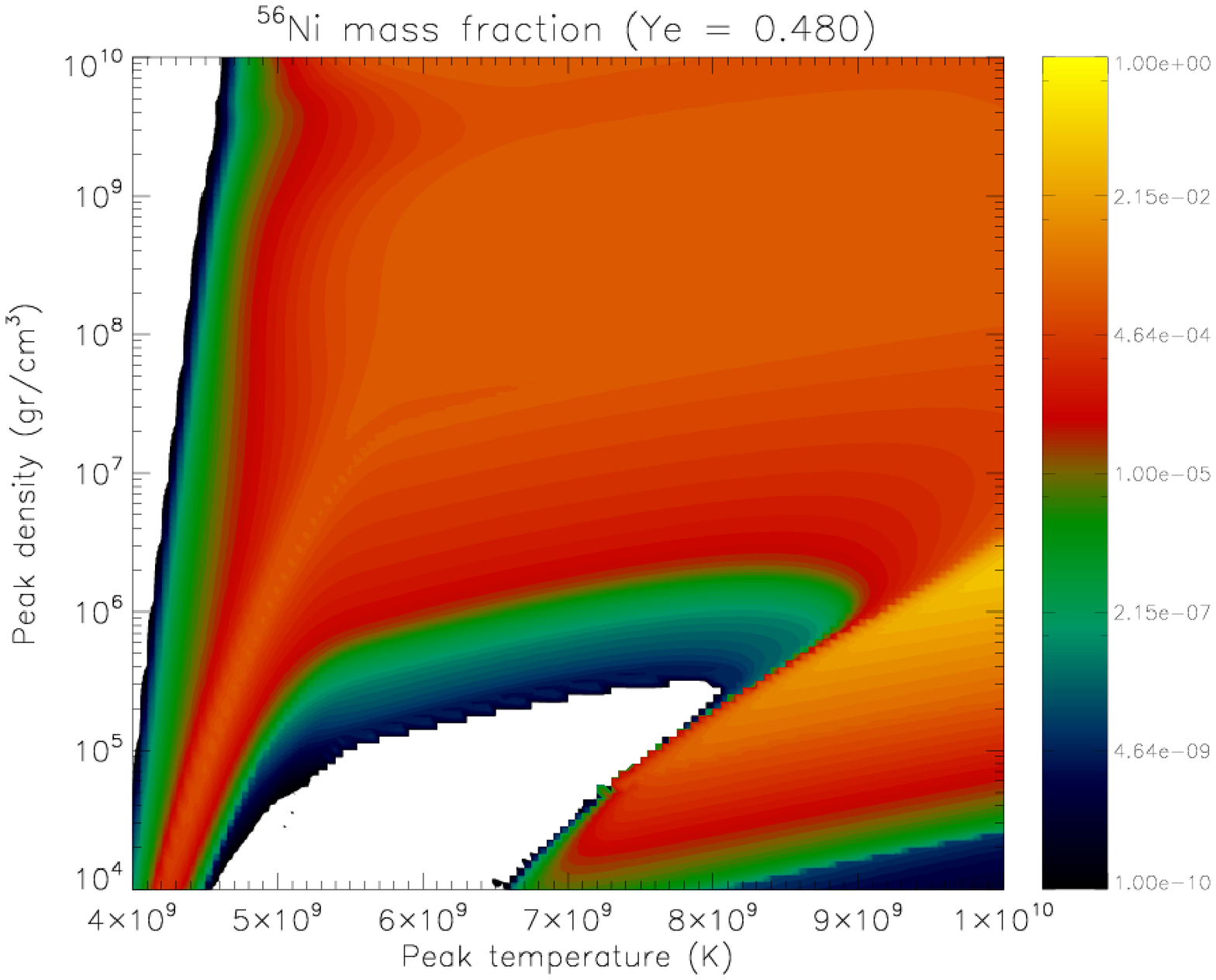}
\end{center}
\end{minipage}
\begin{minipage}{0.33\textwidth}
\begin{center}
\includegraphics[width=\textwidth]{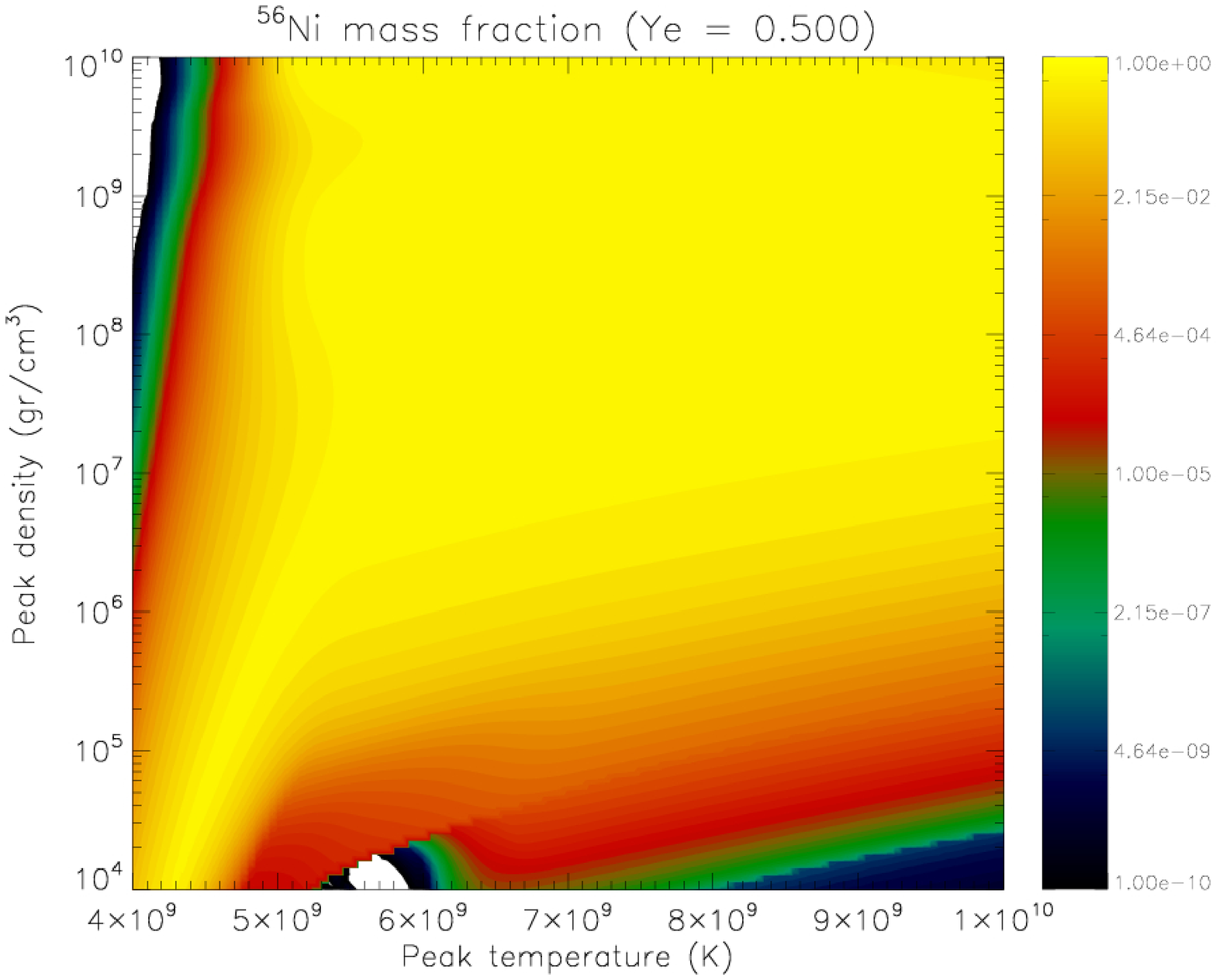}
\end{center}
\end{minipage}
\begin{minipage}{0.33\textwidth}
\begin{center}
\includegraphics[width=\textwidth]{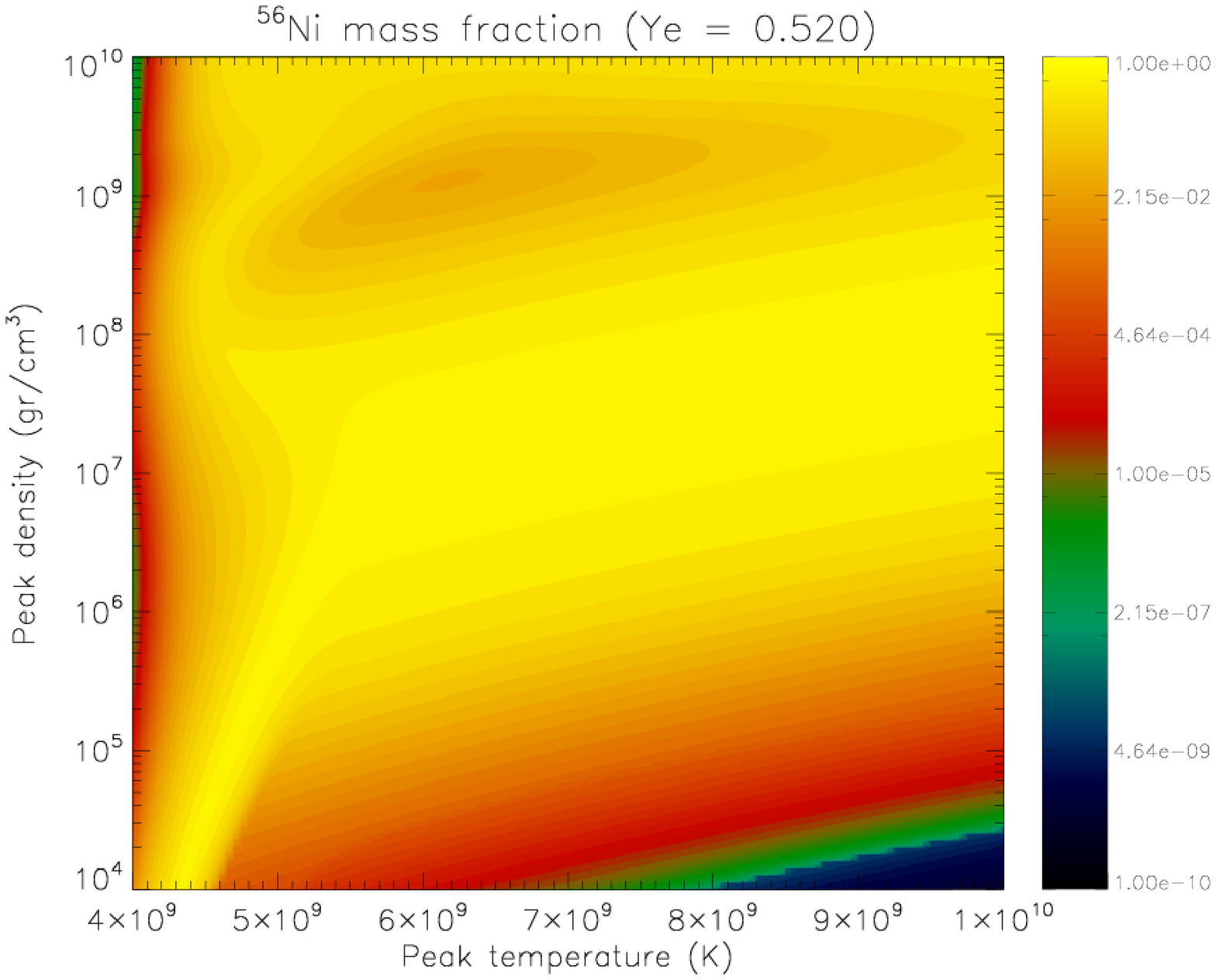}
\end{center}
\end{minipage}
\caption{$\ux{56}{Ni}$ contour plots for adiabatic expansions with
initial electron fraction $\mr{Y_{e}}=0.48$ (left),
$\mr{Y_{e}}=0.50$ (middle) and $\mr{Y_{e}}=0.52$ (right).}
\label{fig:ni56_ye}
\end{figure}
\end{center}

The yields for $\ux{44}{Ti}$ and $\ux{56}{Ni}$ are sensitive to the
initial value of the electron fraction $\mr{Y_{e}}$, which we study
for the the first time. In general, major flows tend to focus along
isotopes whose proton to nucleon ratio is equal to the current value
of \mr{Y_{e}}. Thus, major flows will go through $\ux{44}{Ti}$ and
$\ux{56}{Ni}$ for \mr{Y_{e}=0.5}, because the number of protons equals
the number of neutrons for both isotopes.

As is well known, $\ux{56}{Ni}$ dominates the final compositions
when \mr{Y_{e}=0.5} (Fig.\ \ref{fig:ni56_ye}), because of its
relatively high binding energy.  For \mr{Y_{e}<0.5} the major flows
proceed towards more neutron rich, heavier nuclei and $\ux{56}{Ni}$
does not dominate the final composition. For \mr{Y_{e}>0.5}, instead
of being dominated by the Fe-peak nuclei with the largest binding
energy per nucleon that have a proton to nucleon ratio close to the
prescribed electron fraction, proton-rich material minimizes its
Helmholtz free energy by being mainly composed of $\ux{56}{Ni}$ and
free protons \cite{ivo}.

\begin{figure}[ht]
\begin{minipage}{0.5\textwidth}
\begin{center}
\includegraphics[width=\textwidth]{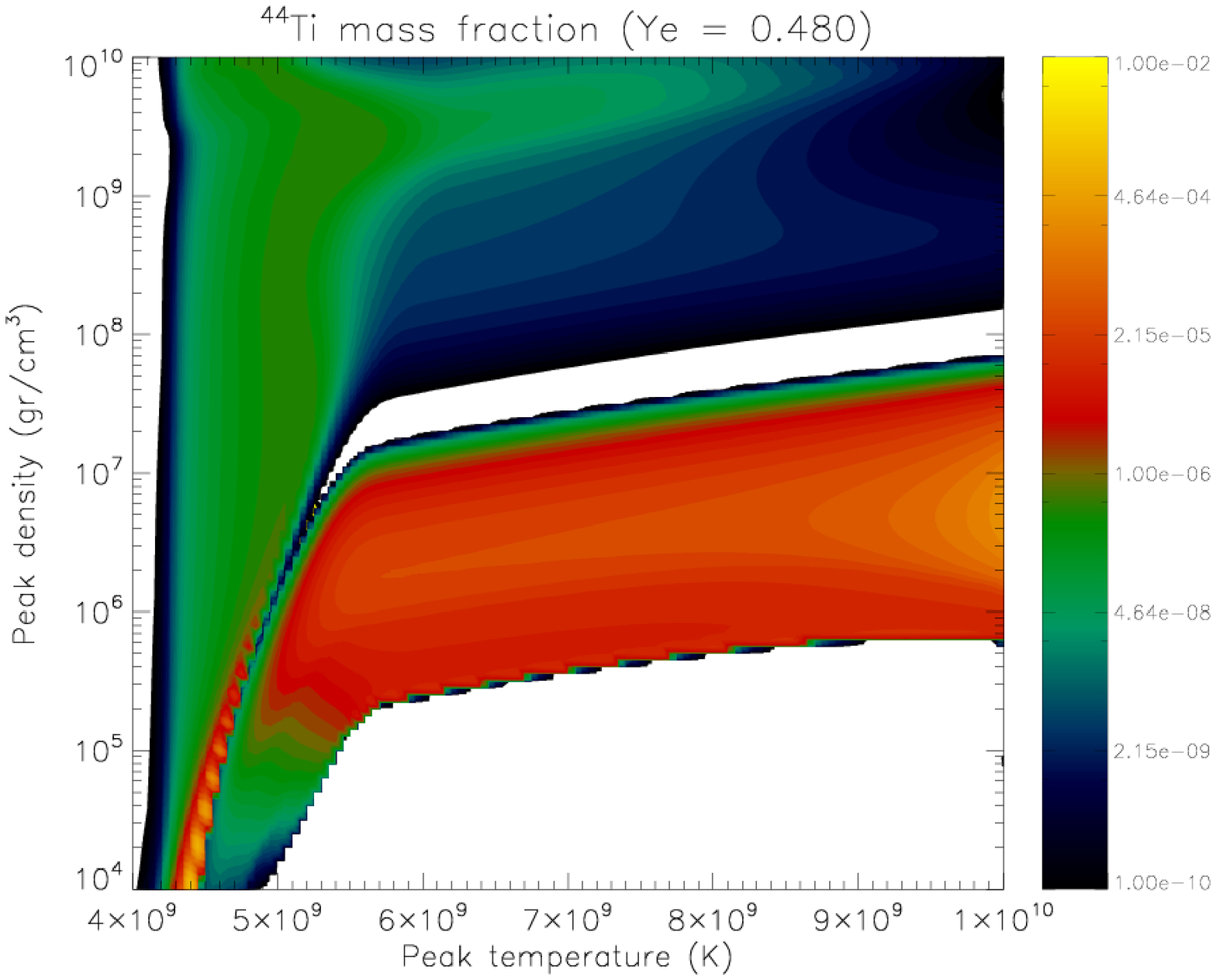}
\end{center}
\end{minipage}
\begin{minipage}{0.5\textwidth}
\begin{center}
\includegraphics[width=\textwidth]{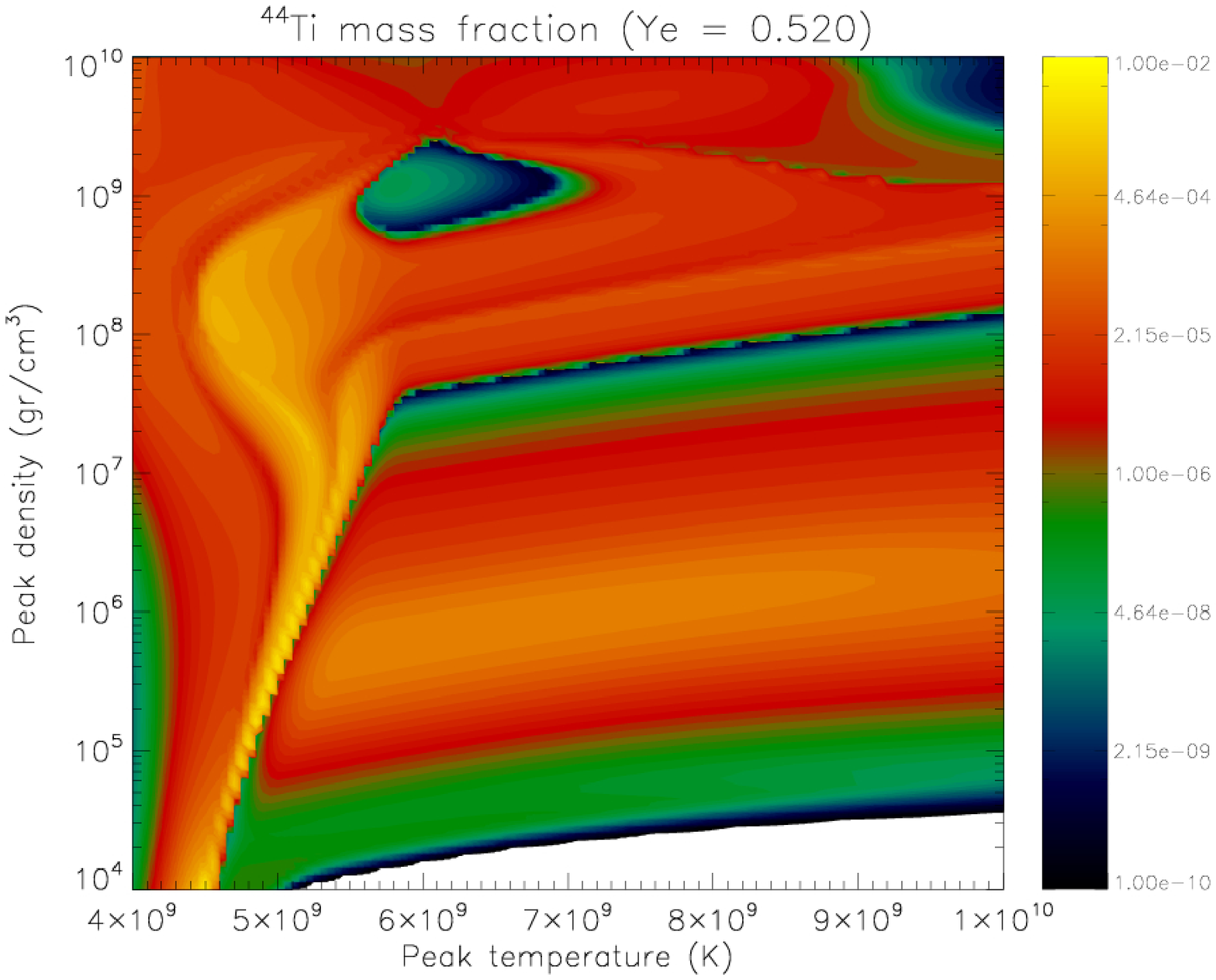}
\end{center}
\end{minipage}
\caption{$\ux{44}{Ti}$ contour plots for $\mr{Y_{e}}=0.48$ (left)
and $\mr{Y_{e}}=0.52$ (right). Power-law expansions are used for
both contours.} \label{fig:ti44_ye_0p48_0p52}
\end{figure}

Fig.\ \ref{fig:ti44_ye_0p48_0p52} shows the $\ux{44}{Ti}$ yields for
two different values of \mr{Y_e}. For \mr{Y_{e}<0.5}, it is bypassed
by the major flows and hence underproduced. For \mr{Y_{e}>0.5},
although it is bypassed again, weak interactions with a halflife
shorter to the expansion timescale restore it mass fraction. These
reactions include primarily
\mr{\ux{44}{V}(e^{-},\nu_{e})\ux{44}{Ti}} and
\mr{\ux{42}{Ti}(e^{-},\nu_{e})\ux{42}{Sc}}. Additionally,
\mr{\ux{39}{Ca}(e^{-},\nu_{e})\ux{39}{K}} and
\mr{\ux{43}{Ti}(e^{-},\nu_{e})\ux{43}{Sc}} seem to affect the yield
for \ux{44}{Ti}.


\end{document}